\documentclass[12pt]{iopart}

\usepackage{iopams}
\usepackage{epsfig}
\usepackage{cite}

\newcommand{\be}{\begin{equation}}
\newcommand{\ee}{\end{equation}}
\newcommand{\bea}{\begin{eqnarray}}
\newcommand{\eea}{\end{eqnarray}}
\newcommand{\beas}{\begin{eqnarray*}}
\newcommand{\eeas}{\end{eqnarray*}}

\begin{document}

\title{Numerical Solitons of Generalized Korteweg-de Vries Equations}

\author{Houde Han and Zhenli Xu}

\address{Department of Mathematics, University of Science and Technology of China, Hefei, Anhui, 230026, P. R. China}
\ead{hhan@math.tsinghua.edu.cn(H. Han); xuzl@ustc.edu(Z. Xu)}

\begin{abstract}
We propose a numerical method for finding solitary wave solutions
of generalized Korteweg-de Vries equations by solving the
 nonlinear eigenvalue problem on an unbounded domain. The artificial boundary conditions are obtained
 to make the domain finite. We specially discuss the soliton solutions of the $K(m, n)$ equation
 and KdV-$K(m,n)$ equation. Furthermore for the mixed models of linear and nonlinear dispersion,
  the collision behaviors of soliton-soliton and soliton-antisoliton are observed.
\end{abstract}

\pacs{52.35.Sb, 05.45.Yv, 47.35.+i}

\maketitle
\section{Introduction}

In this paper, we propose a numerical method for finding solitary
wave solutions, which we call numerical solitons,
 of generalized Korteweg-de Vries (GKdV) equations.
The method is simple and effective to find numerical solitons and
can thus be used to study the nonlinear mechanism of nonlinear
evolution equations, especially in equations for which it is
difficult to obtain soliton solutions by analytical tools. A
sequence of GKdV equations which are not necessarily integrable
are shown to admit soliton solutions by our method. In general,
the GKdV equations are of the form
\begin{equation}\label{gkdv}
u_t+G(u, u_x)_x+H(u)_{xxx}=0,
\end{equation}
where $G$ and $H$ are given functions. The $K(m,n)$ equation with
$m,n\ge 1$,
\begin{equation}\label{kmn}
u_t+(u^m)_x+(u^n)_{xxx}=0,
\end{equation}
is a special case of the GKdV equation (\ref{gkdv}).

When $m=2,~n=1$, it is the first important model equation, the
Korteweg-de Vries (KdV) equation,
\begin{equation}\label{kdv}
u_t+(u^2)_x+u_{xxx}=0,
\end{equation}
which is integrable and admits $sech^2$ soliton solutions. For the
KdV equation (\ref{kdv}), the linear dispersion and the nonlinear
convection work against each other and exactly balance to result
in a stable solution. Since Zabusky and Kruskal \cite{ZK:PRL:65}
numerically studied the elastic collisions of KdV solitons, the
research of solitons has attracted many workers' attentions and
interests.

For $n>1$, the $K(m,n)$ equation(\ref{kmn}) is fully nonlinear
\cite{RH:PRL:93}. Their soliton solutions have compact support and
called compactons. For $m=n$, they assume a very simple
form\cite{RH:PRL:93}, namely
\begin{equation}
u=\{\frac{2 \lambda n}{n+1}cos^2[\frac{n-1}{2n}(x-\lambda
t)]\}^{\frac{1}{n-1}},~~ \mathrm{for}~~ |x-\lambda t|\le
\frac{n\pi}{n-1},
\end{equation}
and zero otherwise, where $\lambda$ is a constant which represents
the velocity
 of the travelling wave. The $K(m,n)$ equation is not integrable \cite{RH:PRL:93,R:PLA:97}; however, the
 collisions of the compactons are almost elastic by numerical experiments \cite{RH:PRL:93}.
 This suggests that the elastic collisions are shared by many nonlinear equations, even though they
 are not integrable. Hence, it seems of great interest to observe more equations to further study
 their nonlinear mechanisms.

 However, the analytic ways are limited for many equations, due to the complexity of the nonlinear properties.
  The aim of the proposed numerical method is to obtain soliton solutions for these equations.
 Our approach is to transform the nonlinear eigenvalue problems on the unbounded domain deduced from
 the travelling wave solutions of the equations into bounded problems by imposing an artificial
 boundary condition. These problems can then be solved numerically in an efficient way.

 The rest of the paper is organized as follows. In section 2, we discuss the numerical method.
 In section 3, some application examples are proposed by using the method. Conclusions are
 given in section 4.

\section{The method}

\subsection{$K(m,n)$ equations}

Although the method is suitable for wider classes of nonlinear
evolution equations, here we introduce the idea by considering the
solitary wave solutions of $K(m,n)$ equation (\ref{kmn}) for
$m,n\ge1$. Suppose that
\begin{equation}
u(x, t)=f(\xi)=f(x-\lambda_1 t)
\end{equation}
is a travelling wave solution of the $K(m,n)$ equation. It is easy
to see that $\lambda_1$ and $f(\xi)$ satisfy the nonlinear
eigenvalue problem \bea
&&-\lambda_1f+f^m+(f^n)_{\xi\xi}=0,\label{eq06}\\
&&f(\xi)\rightarrow 0~~~~\mathrm{as}~\xi\rightarrow \pm\infty.
\eea In general, the spectrum is continuous, and if $f(\xi)$ is an
eigenfunction corresponding to $\lambda_1$, then so is
$f(\xi-\xi_0)$ for any $\xi_0\in R$. Hence, it is very difficult
to solve (6), (7) numerically.

If, however, we prescribe the amplitude of $f$ and the location of
its maximum, for example by setting \bea
&&f(0)=1,\\
&&f'(0)=0, \eea then (6)-(9) represents a nonlinear boundary-value
problem on $(-\infty,0]$ and $[0,\infty)$. Moreover, if
$\lambda_1$ and $f(\xi)$ is a solution of (6)-(9), then
\begin{equation}u(x,t)=Af(B(A)(x-\lambda(A)t)),\label{eq10}\end{equation}
is a solitary wave solution of equation $K(m,n)$, with amplitude
$A>0$, and functions $B(A)$ and $\lambda(A)$ to be determined. Let
$\eta=B(A)(x-\lambda(A)t)$. Substituting (\ref{eq10}) into
equation (\ref{kmn}), and using equation (\ref{eq06}), we obtain
\be \begin{array}{l}
-A\lambda(A)f+A^mf^m+A^nB(A)^2(f^n)_{\eta\eta}\\
~~~~~~=(-A\lambda(A)+A^m\lambda_1)f+(A^nB(A)^2-A^m)(f^n)_{\eta\eta}.
\end{array}\ee
We set \be B(A)=A^{\frac{m-n}{2}},\ee and \be
\lambda(A)=\lambda_1A^m.\ee We know that $u(x,t)$ is a solitary
wave solution of $K(m,n)$ with amplitude $A$ and velocity
$\lambda_1A^m$. Therefore, we only need to find $\lambda_1$ and
$f(\xi)$, and this allows us to find the other solitary wave
solutions of $K(m,n)$.

For the numerical solution of the problem (6)-(9) on the unbounded
domain $(-\infty, +\infty)$, we only need to consider the problem
on the domain $[0,\infty)$, for the reasons of symmetry. We
introduce an artificial boundary at the point $\xi=b>0$, where $b$
is large enough. We use the boundary condition $f(b)=0$ to reduce
the problem to a problem on the bounded interval $[0,b]$ for
$m>1,~n>1$:
$$\mathrm{Find}~ \lambda_1\in R ~\mathrm{and}~ f(\xi), ~\mathrm{such~ that}$$
\bea
&&-\lambda_1f+f^m+(f^n)_{\xi\xi}=0, ~0<\xi<b\\
&&f(0)=1,\\
&&f'(0)=0,\\
&&f(b)=0. \eea

We solve the problem numerically by a finite difference method.
Let $h=b/N$ with a positive integer $N$,
$\xi_i=ih,~i=0,1,\cdots,N$. Then the interval $[0,b]$ is divided
into $N$ subintervals by
$$0=\xi_0<\xi_1<\cdots<\xi_N=b.$$
Using the second-order central difference, we obtain \be
-\lambda_1f_i+f_i^m+\frac{f_{i+1}^n-2f_i^n+f_{i-1}^n}{h^2}=0,
~~i=0,1,\cdots,N-1.\ee Due to the boundary conditions we have \be
f_{-1}=f_1,~~f_0=1, ~~f_N=0.\ee The problem (18) with boundary
conditions (19) is a nonlinear algebraic system which can be
solved by the Newton iteration method.

\subsection{MKdV-$K(m,n)$ equation}
The second model we consider here is the mKdV-$K(m,n)$ equation
\be u_t+(u^m)_x+(u+u^n)_{xxx}=0.\ee We also assume the travelling
wave solution \be u(x,t)=f(\xi)=f(x-\lambda t).\ee Then we have
\begin{equation}
-\lambda f_{\xi}+(f^m)_{\xi}+(f+f^n)_{\xi\xi\xi}=0,
\end{equation}
which can be integrated once to yield
\begin{equation}\label{once}
-\lambda f+f^m+(f+f^n)_{\xi\xi}=C,
\end{equation}
where $C$ is an arbitrary constant.   Since
 we are interested in a localized solution, we know that:

 (i). $f(\xi)\rightarrow 0,$ as $\xi\rightarrow\pm \infty$; therefore we have $C=0$ in equation (23);

(ii). $f(0)=A$ is an extremum, i.e., $f'(0)=0$.

Similarly, we obtain the following nonlinear problem on
$[0,\infty)$: \bea
&&-\lambda f+f^m+(f+f^n)_{\xi\xi}=0,\\
&&f(\xi)\rightarrow 0~~~~\mathrm{as}~\xi\rightarrow +\infty,\\
&&f(0)=A,\\
&&f'(0)=0. \eea We introduce the artificial boundary point
$\xi=b>0$, with $b$ large enough, and $m,n>1$. On the interval
$[b,\infty)$, the equation (24) can be linearized to \be -\lambda
f+f_{\xi\xi}=0, ~~~\xi\in[b,\infty),\ee which has two linearly
independent solutions,
$$e^{\sqrt{\lambda}\xi}~~~\mathrm{and}~~~e^{-\sqrt{\lambda}\xi}.$$
By the condition (25), $f(\xi)=Ce^{-\sqrt{\lambda}\xi}$ with
$\lambda>0,$ and
$$f'(\xi)=-\sqrt{\lambda}Ce^{-\sqrt{\lambda}\xi}=-\sqrt{\lambda}f(\xi).$$
Then at $\xi=b$ we have the artificial boundary condition \be
f'(b)=-\sqrt{\lambda}f(b).\ee Thus we have reduced the problem to
a problem on the bounded interval $[0,b]$: \bea
&&-\lambda f+f^m+(f+f^n)_{\xi\xi}=0,~~0<\xi<b\\
&&f(0)=A,\\
&&f'(0)=0,\\
&&f'(b)=-\sqrt{\lambda}f(b), \eea which can again be solved by the
finite difference method.

\section{Numerical examples}

{\bf Example 1.} Use the proposed method in Section II, numerical
solitary wave solutions of the GKdV equation can be obtained.
 We first consider the $K(2,2)$ and $K(3,3)$ equations to test the efficiency of the method.

 \begin{center}
 (a)\epsfig{file=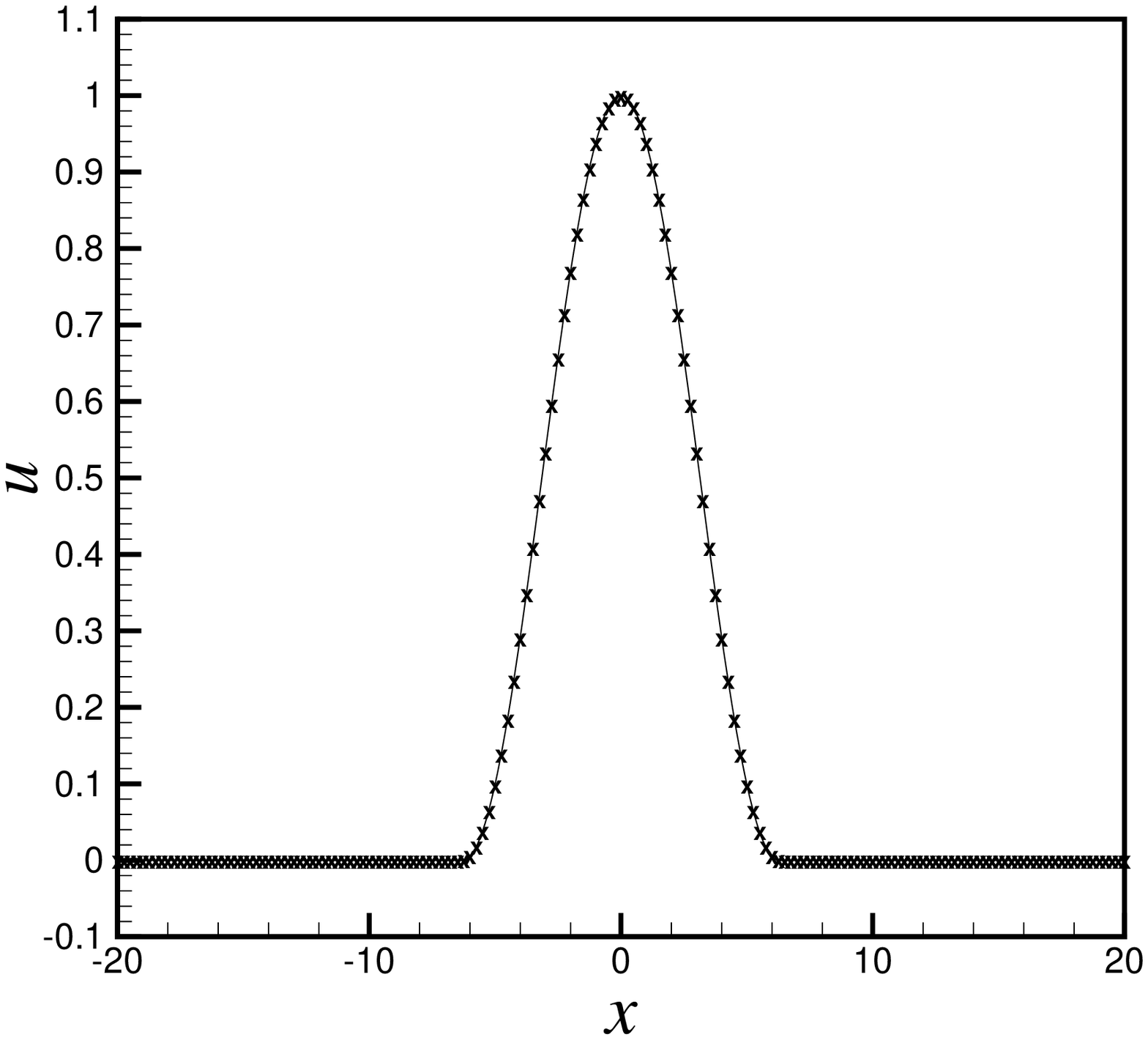, width=6cm} (b)\epsfig{file=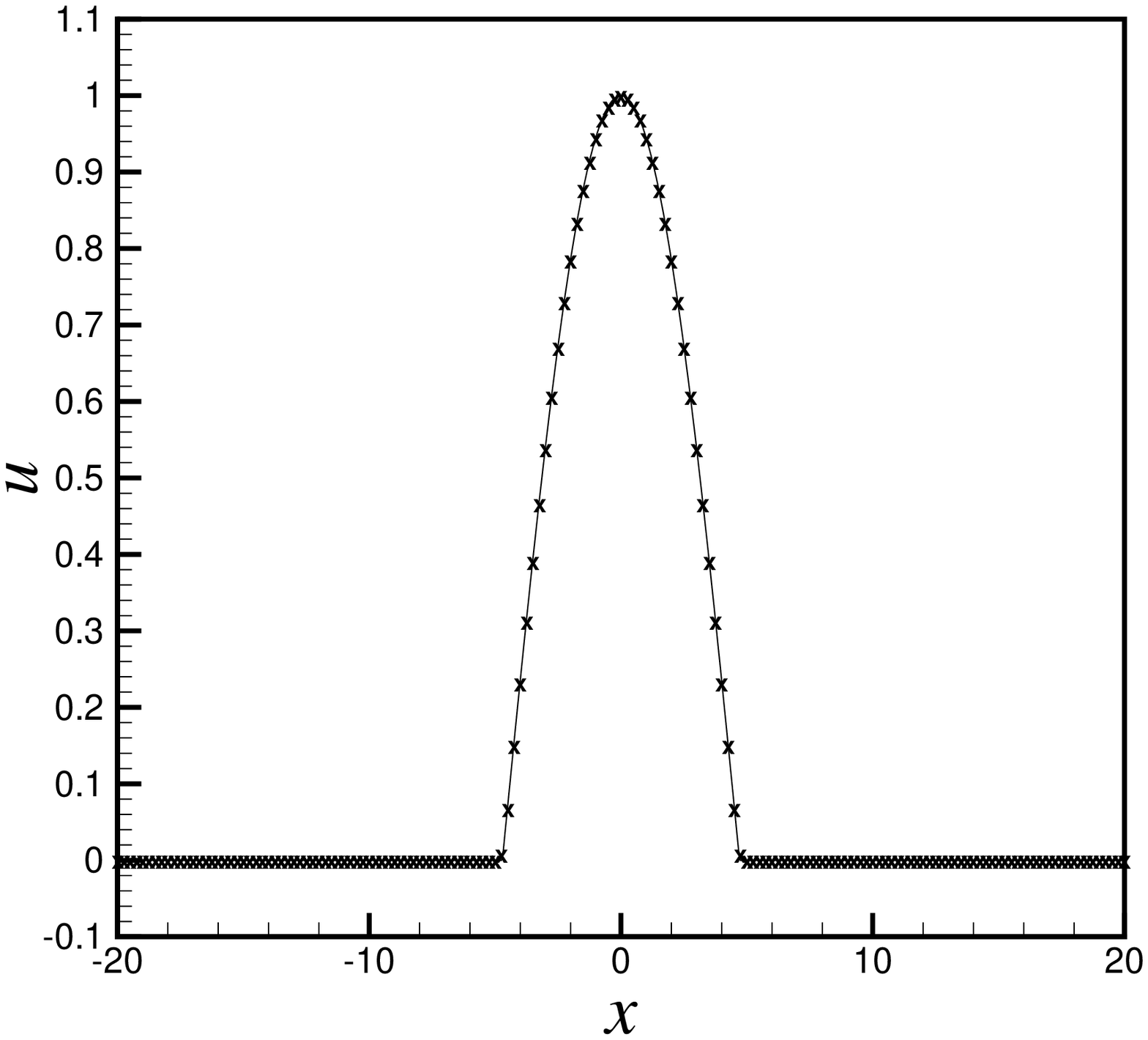, width=6cm}\\
{\scriptsize Fig. 1. Numerical solitons. Solid line is the exact
solution.
 (a) $K(2,2)$; (b) $K(3,3)$.}
\end{center}

 In this case, the Newton iteration maybe does not converge or the solution shows blowup phenomena
with the grid refinement. To avoid these phenomena, a low pass
filter has been used in iterate steps. Fig. 1 (a) and (b) show the
numerical results with $h=0.25$. We see the numerical results are
well agreed with the exact solutions.

{\bf Example 2.} We show the numerical soliton to the KdV-$K(2,2)$
equation
\begin{equation}\label{kdvk22} u_t+(2u^2)_x+(u+u^2)_{xxx}=0,\end{equation}
which is a combination of the KdV and $K(2,2)$ equations. Unlike
the KdV (\ref{kdv}) and $K(m,n)$ (\ref{kmn}) equations, it is a
mixed dispersive model. Both linear and nonlinear dispersions are
often presented
 in many physical applications, for example the Carmassa-Holm equation \cite{CH:PRL:93} in water
waves. It is interesting to solve them to study the interaction
mechanism of linear and nonlinear structures \cite{FOKAS:PD:95}.

Fig. 2(a) illustrates the shapes of three numerical solitons with
amplitudes $A=2$, 1.5 and 1, respectively. We also perform the
numerical experiments of collisions with the three solitons which
are located at the three isolated centers $x=-60$, -30 and 0.
Here, the fully implicit Crank-Nicolson scheme is used in the
domain $[-80, 80]$ with periodic boundary conditions. Fig. 2(b-d)
shows the evolution of the three numerical solitons. Extensive
tests all indicate that the collisions of the KdV-$K(2, 2)$
equation (\ref{kdvk22}) are almost elastic.

\begin{center}
(a)\epsfig{file=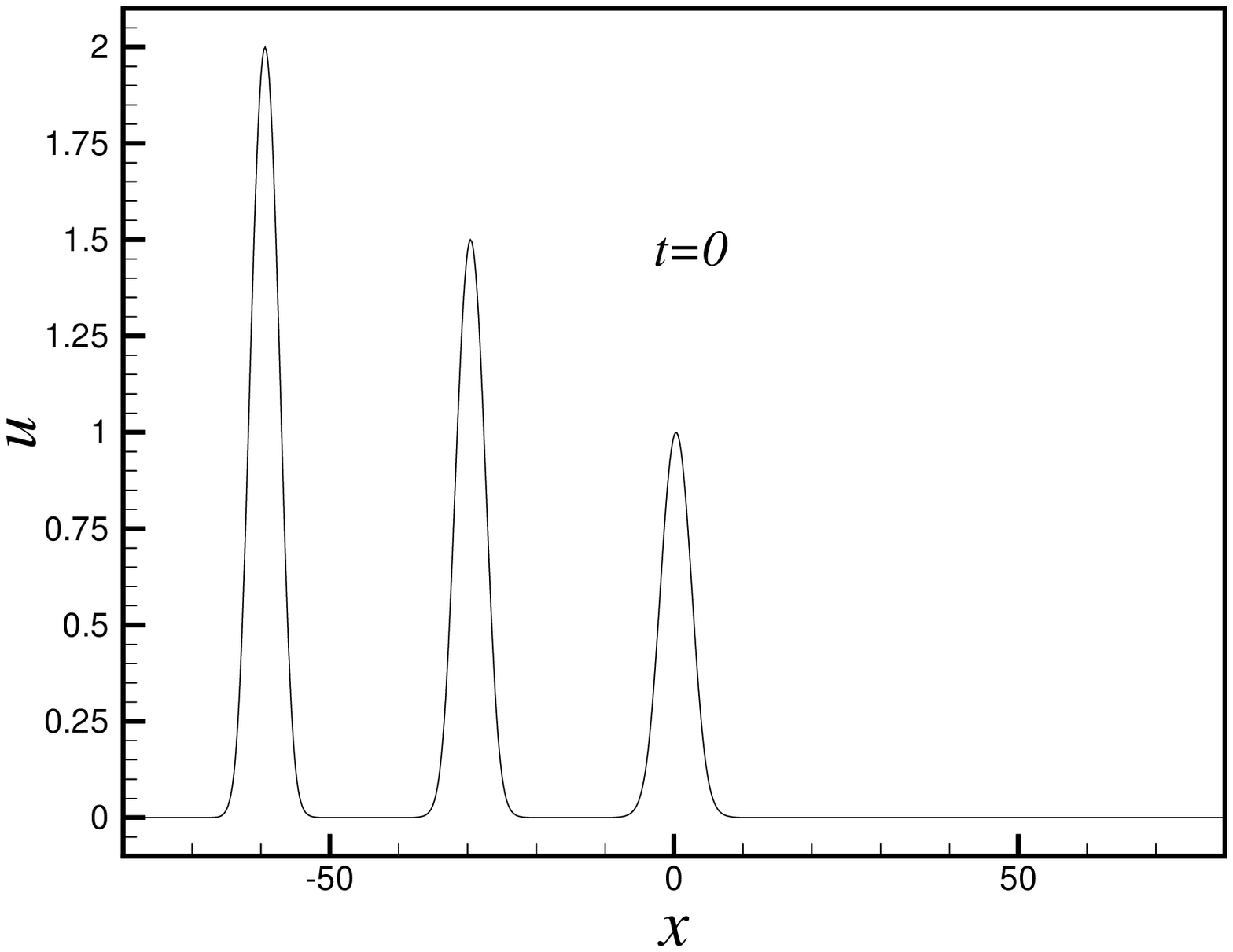, width=6cm} (b)\epsfig{file=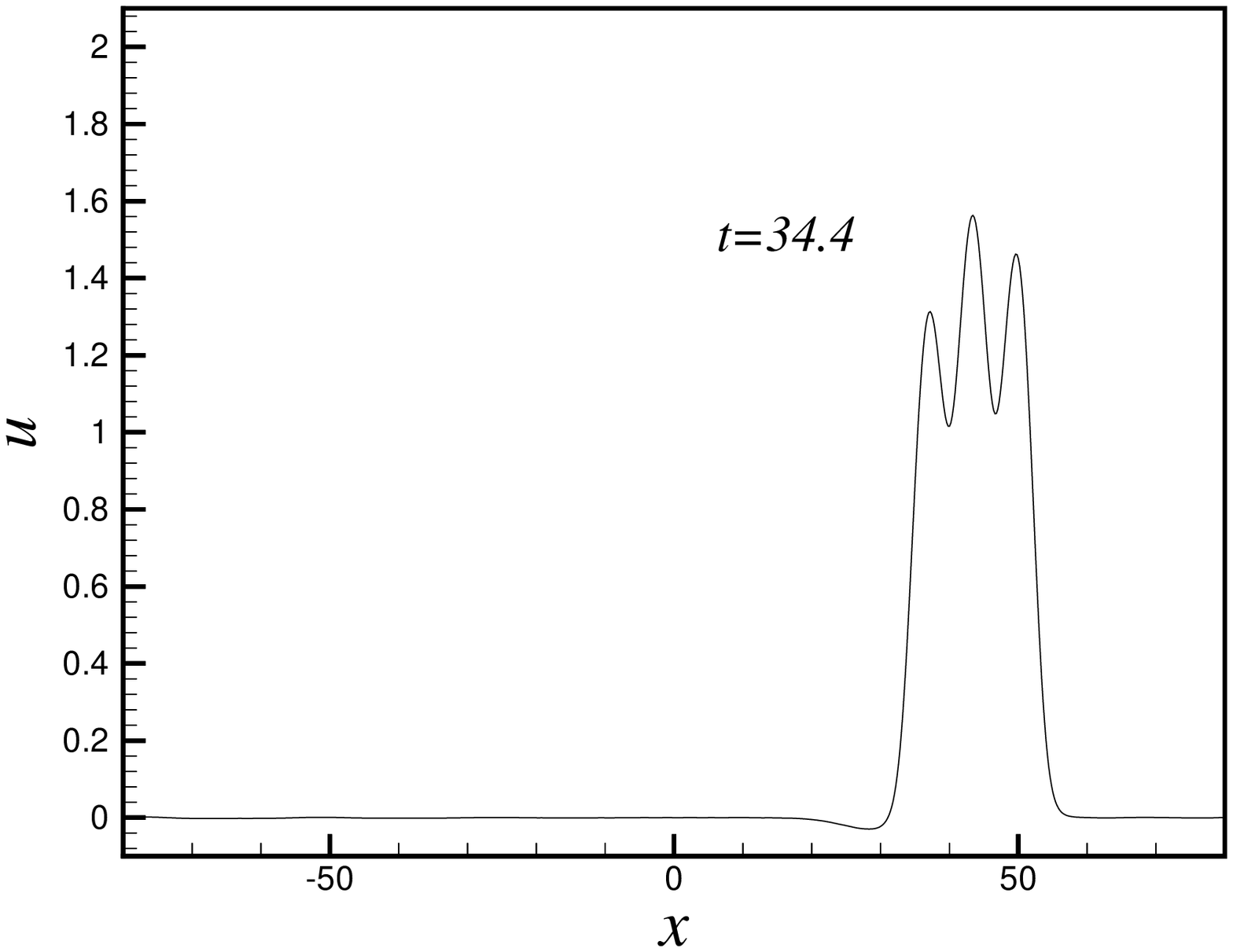, width=6cm}\\
(c)\epsfig{file=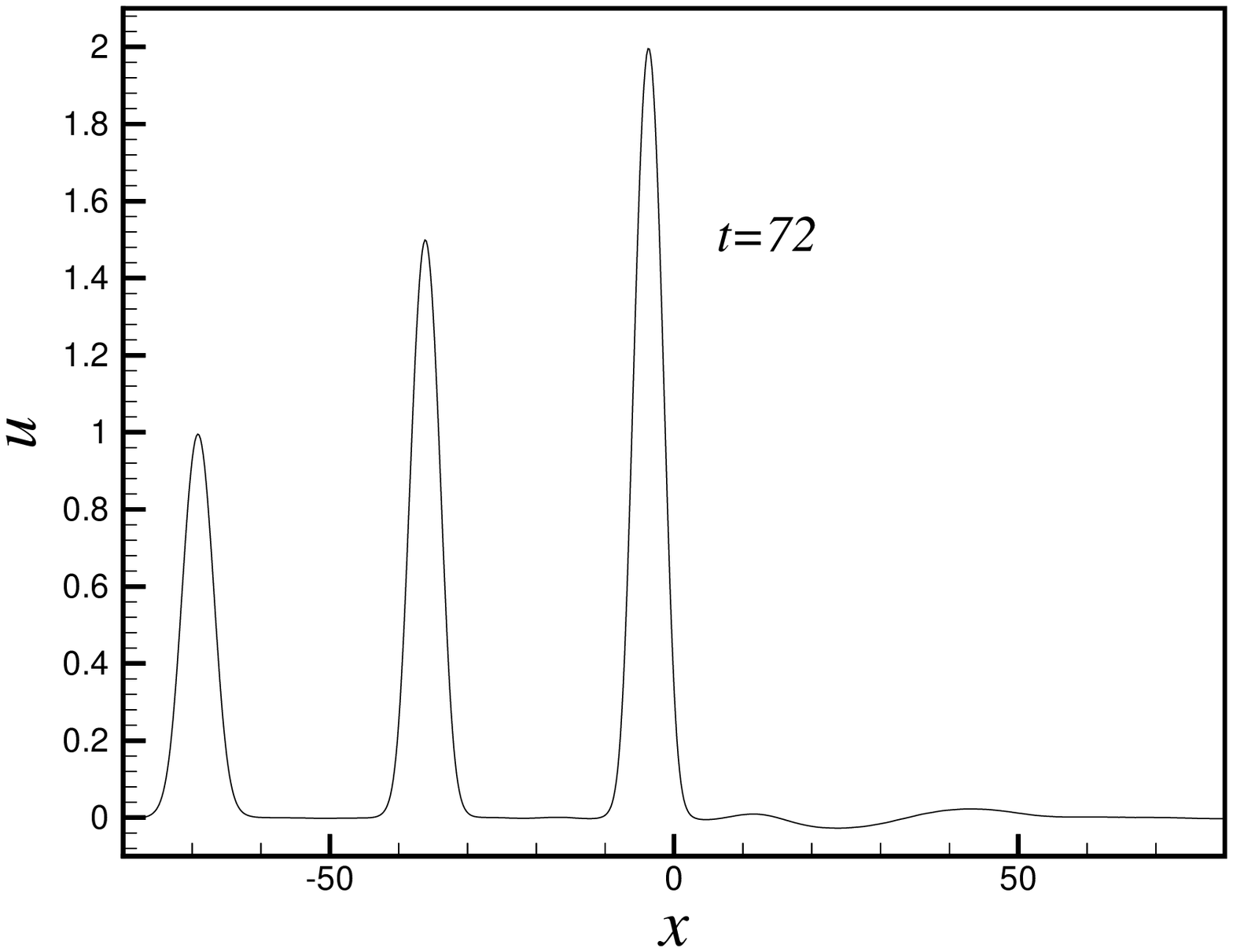, width=6cm} (d)\epsfig{file=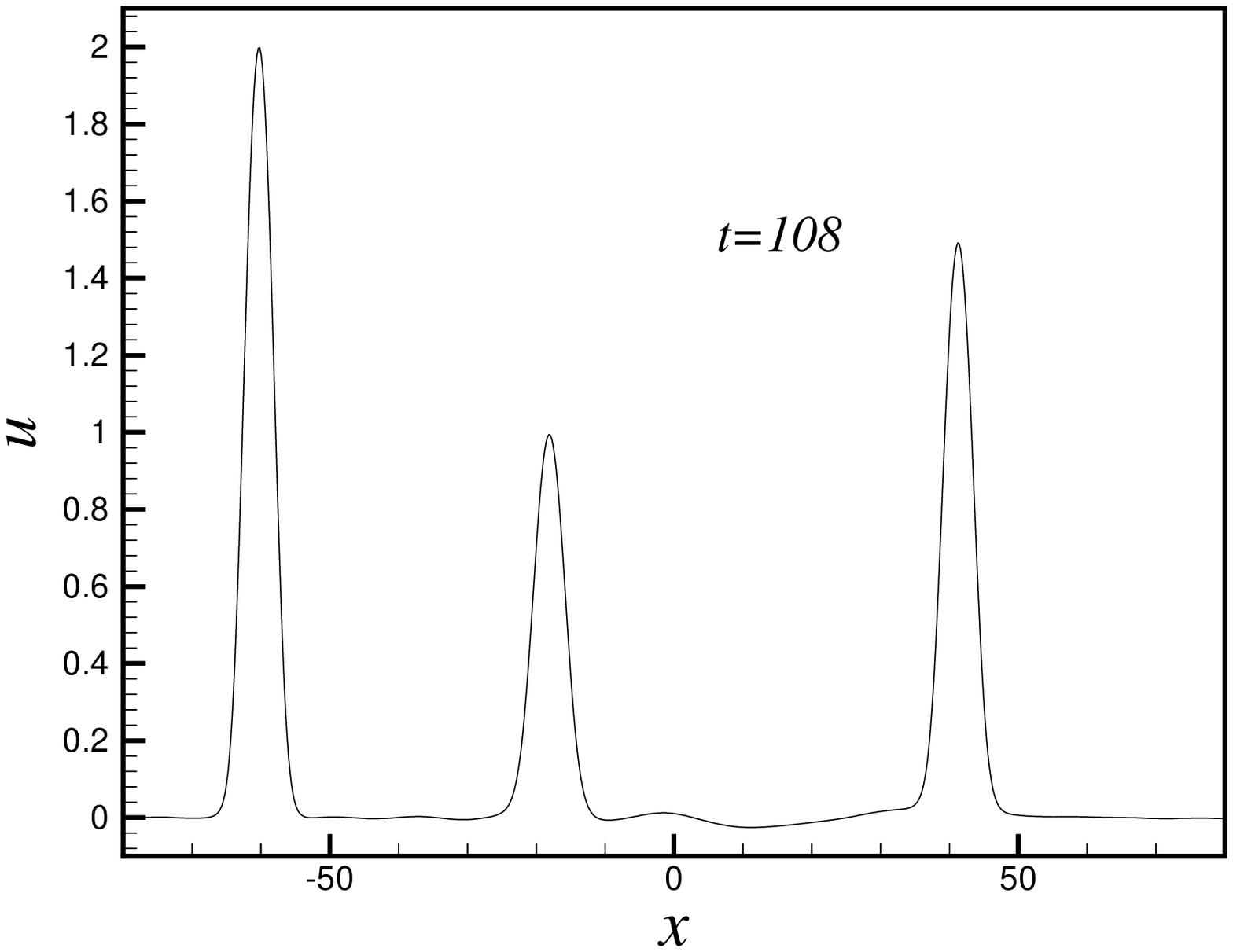, width=6cm} \\
{\scriptsize Fig. 2. The evolution of three numerical solitons.}
\end{center}

Similarly to the behavior of the $K(2, 2)$ compactons, the
collisions of KdV-$K(2, 2)$ numerical solitons also create a small
ripple (see Fig. 3). The amplitudes of the ripple are less than
3\% of the amplitudes of all the solitons. However, their
amplitudes do not almost decrease with the mesh is refined. The
result illustrates that the creation of the collision ripple
probably does not
 arise from the numerical accuracy but from its internal mechanism of the equation.
Though this possibility appears reasonable, it needs to be
confirmed by more proofs since now it has only been deduced from
numerical results.

\begin{center}
(a)\epsfig{file=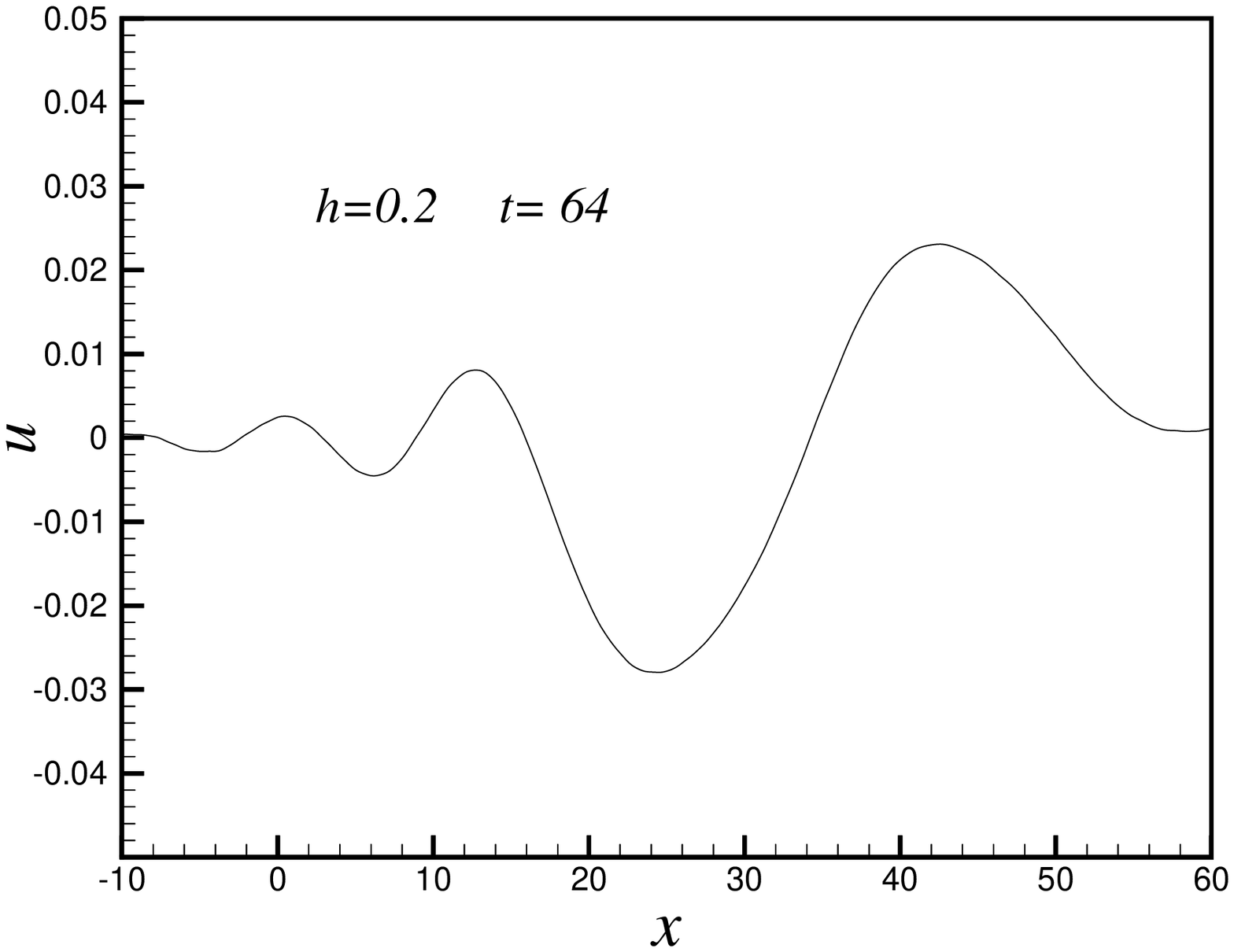, width=6cm} (b)\epsfig{file=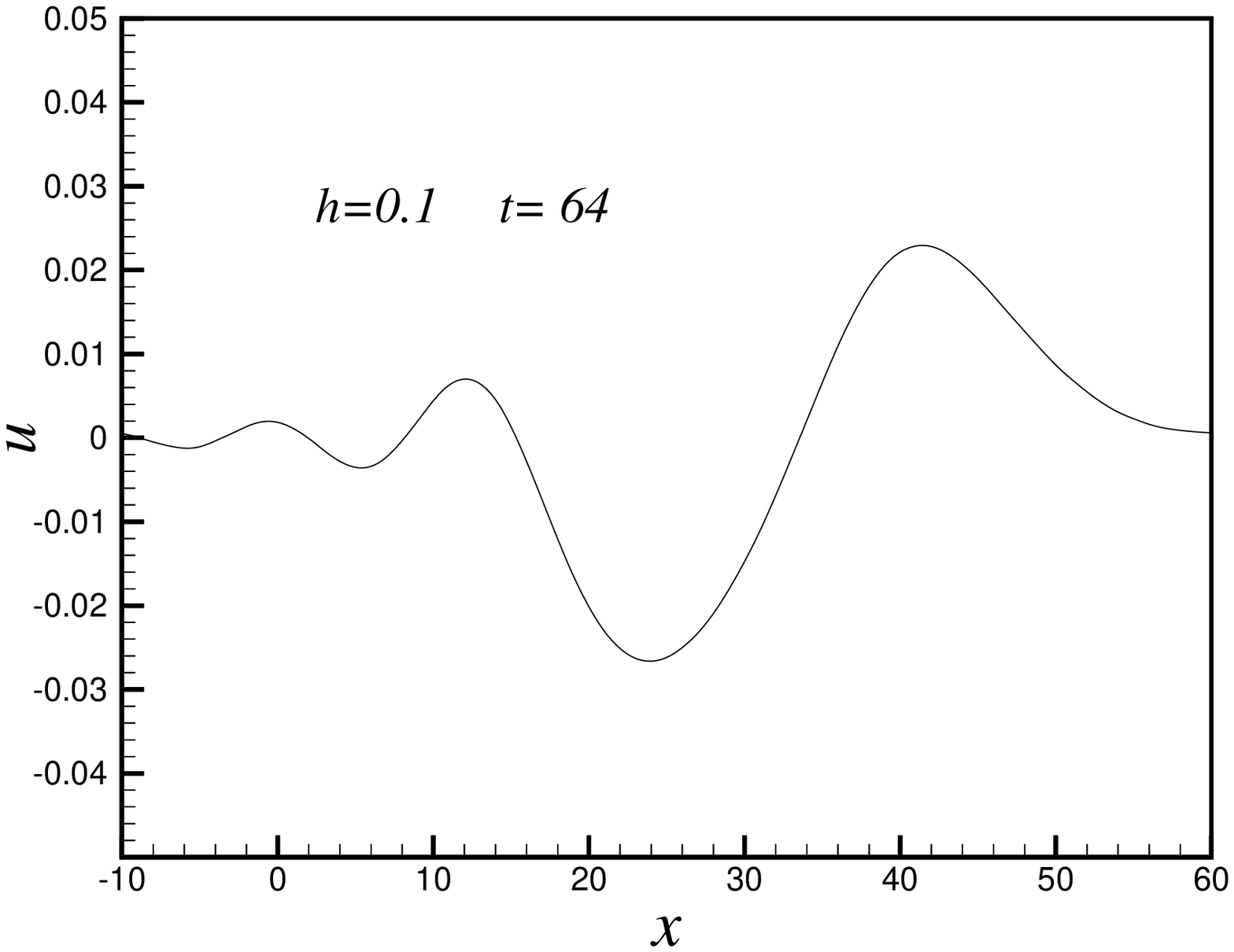, width=6cm} \\
{\scriptsize Fig. 3. The ripples created by the collision of
solitons in different grids.}
\end{center}

{\bf Example 3.} The model equation we considered is called the
mKdV-$K(3,3)$ equation,
\begin{equation}\label{mk3} u_t+(2u^3)_x+(u+u^3)_{xxx}=0,\end{equation}
which is a combination of the modified KdV and $K(3,3)$ equations.
Since all terms of the mKdV-$K(3,3)$ equation (\ref{mk3}) are odd,
its soliton solutions are in pairs by the relationship $(u, -u)$.
They propagate in the same direction with the same speed. The
solitons with negative signs are normally called antisolitons.

The $K(2,2)$ equation also admits anticompactons which propagate
in the opposite direction. However, it is difficult to numerically
simulate compacton-anticompacton collisions because of
instability. It is not clear whether or not the reason of this
instability is due to numerical errors \cite{RH:PRL:93}. Thus, in
order to better understand the reason, it is meaningful to first
study the behavior of soliton-antisoliton collisions.

Fig. 4 shows the evolution of numerical soliton and antisoliton
with amplitudes 2 and 1, respectively. Here we take the spatial
mesh size $h=0.025$ in [-40,40] and periodic boundary conditions.
We can see that their collisions are elastic if the tail
oscillations originate
 from numerical in accuracies. In fact, extensive numerical experiments indicate that the oscillations decrease
when the mesh is refined. This illustrates that the creation of
oscillations are mainly caused by the numerical errors.

\begin{center}
{ (a)\epsfig{file=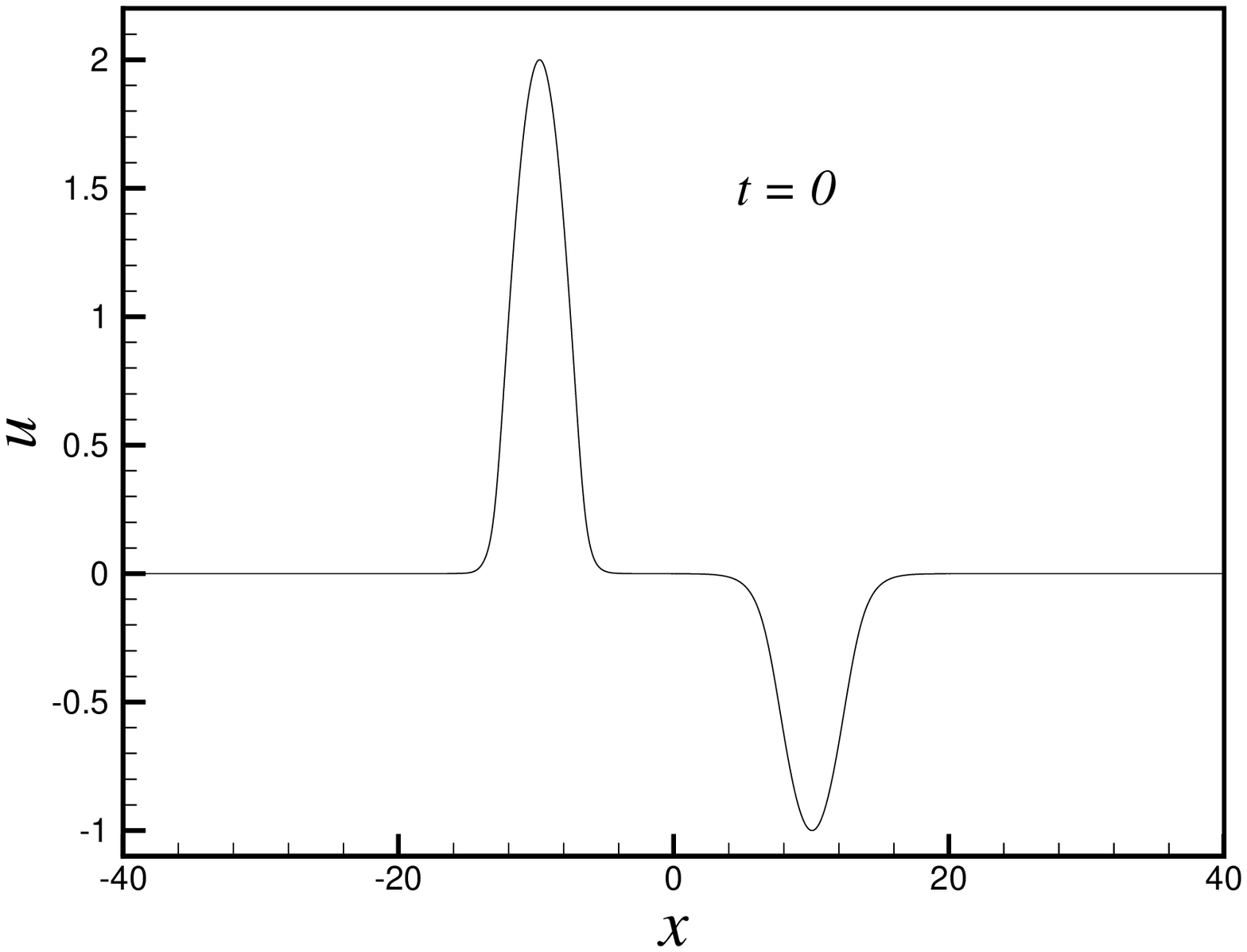, width=6cm} (b)\epsfig{file=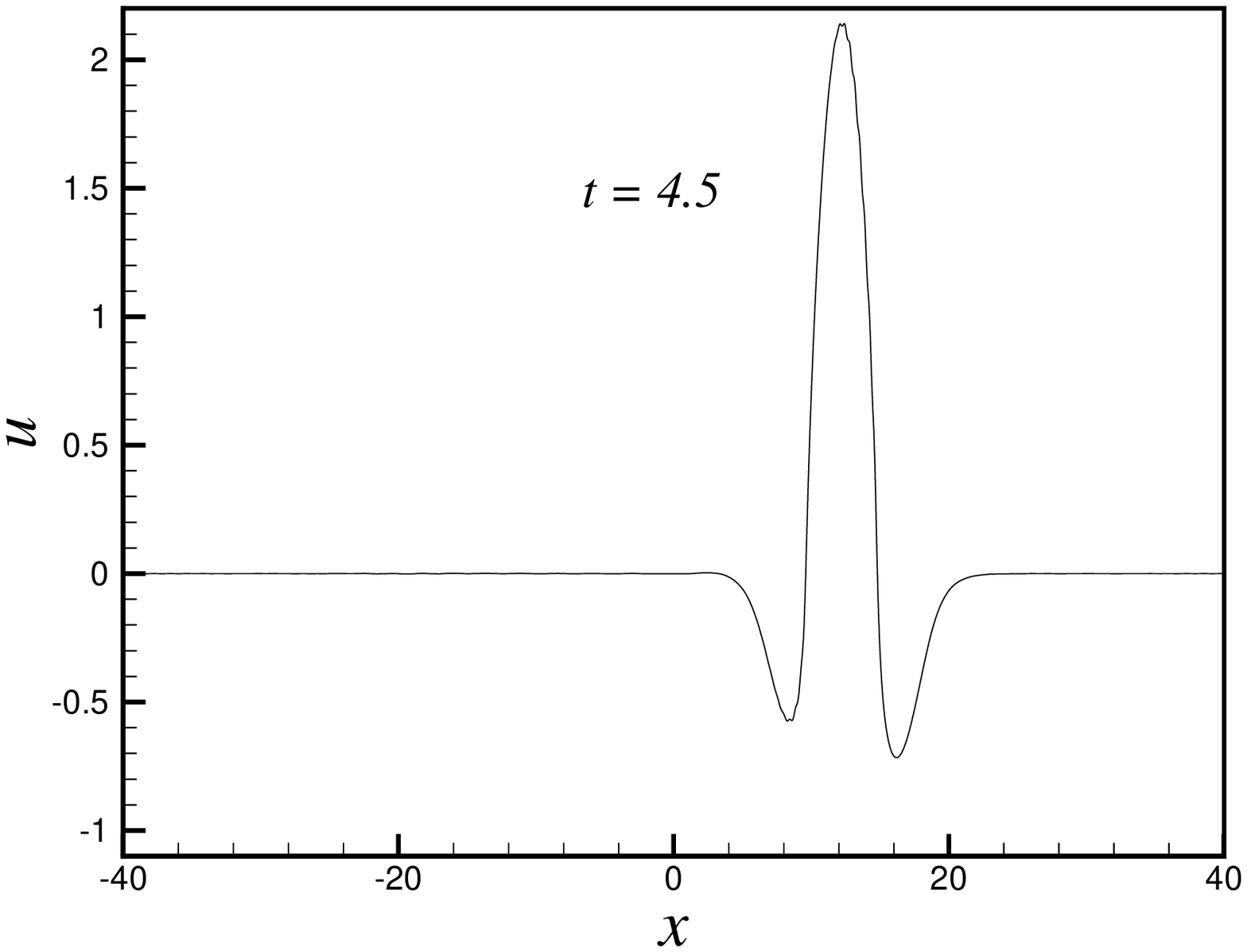, width=6cm}\\
(c)\epsfig{file=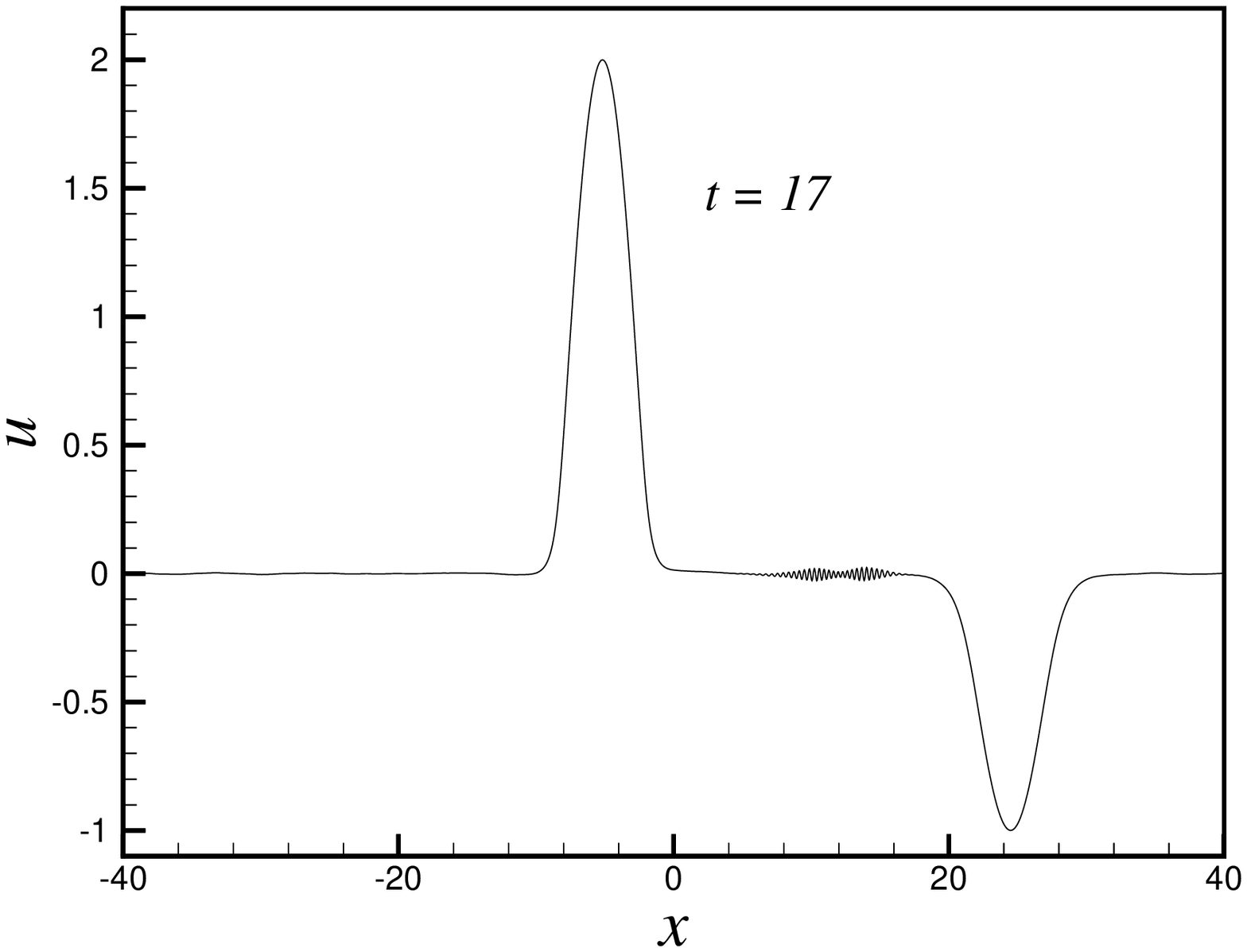, width=6cm} (d)\epsfig{file=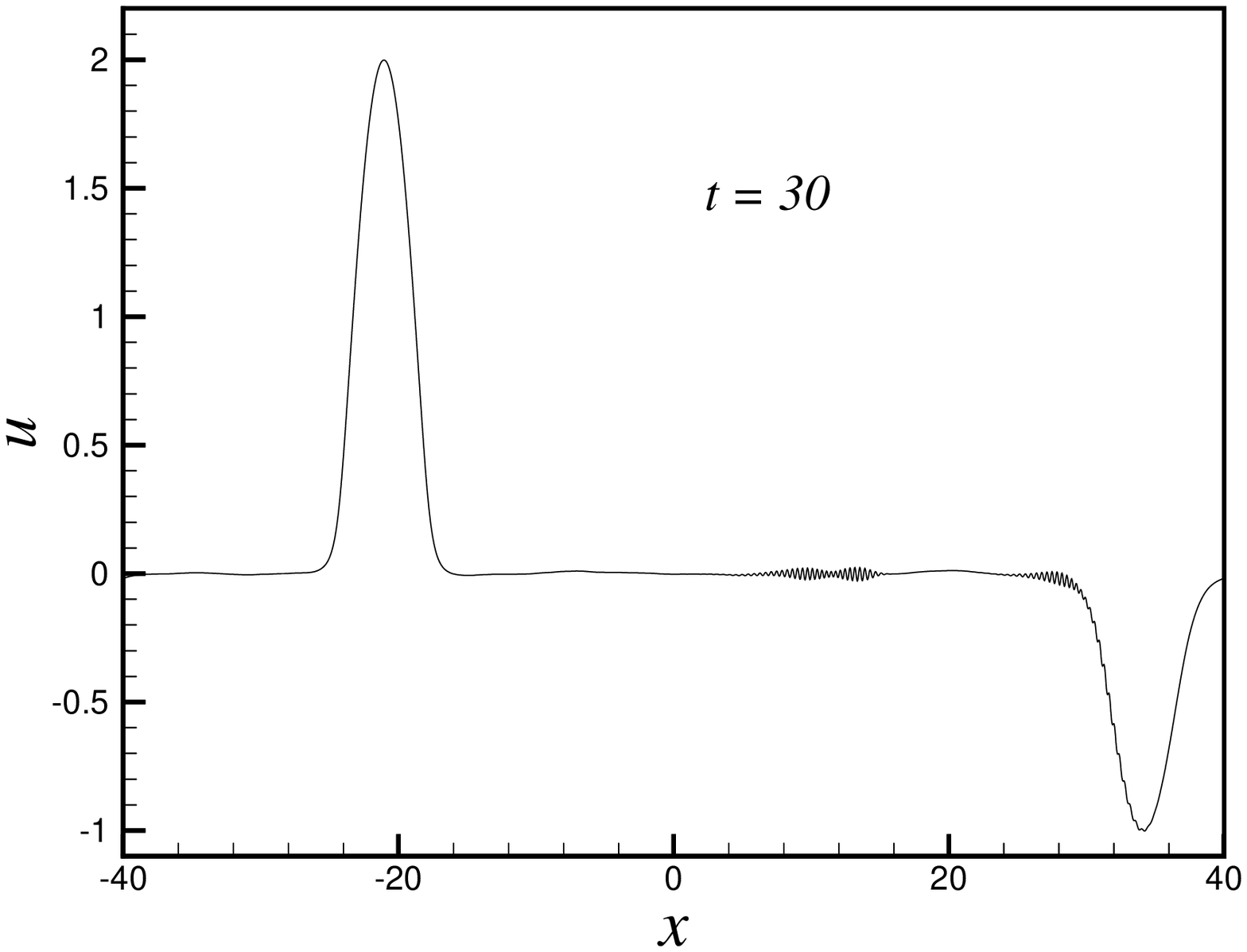, width=6cm}} \\
{\scriptsize Fig. 4. The evolution of soliton-antisoliton
collisions.}
\end{center}

These oscillations are high-frequency waves which are caused by
every collision because of the numerical errors. We see that the
collisions of oscillations and solitons are also elastic (see Fig.
4(c)(d)). However, it is difficult to handle them unless we use a
very fine grid size. This is also the reason why
compacton-anticompacton collisions of the $K(m,n)$ equation are
difficult to be simulated. Based on the above experiments, it is
conjectured that the $K(m, n)$ compacton-anticompacton collisions
are also elastic.

We know the Benjamin-Bona-Mahoney (BBM) equation
\begin{equation}u_t+(u^2)_x-u_{xxt}=0\end{equation}
 as a good example of a nonintegrable model that has also soliton and antisoliton solutions.
Its soliton-soliton collisions are elastic, but numerical tests
illustrate that its soliton-antisoliton
 collisions are inelastic \cite{GG:JCP:90}. Therefore, it is interesting to study what
 governs the behavior of solitons collisions, but we are still a long way from a good understanding of  the wave
 mechanics\cite{LOR:Preprint}.

\section{Conclusion}
In summary, we have reported a method to find numerical solitons
of the GKdV equations which are not necessarily integrable. Using
this method, we have given a basic study of soliton-soliton
collisions of the KdV-$K(2,2)$ equation
 and soliton-antisoliton collisions of the mKdV-$K(3,3)$ equation.
This method and the numerical experiments will contribute to a
future, more extensive study of the theoretical aspects of the
nonlinear dispersive mechanism of evolution equations.

\section*{Acknowledgments}

The authors wish to thank Professor H. Brunner for the valuable
discussions. This work was supported by National Natural Science Foundations of China under Grant
 No. 10471073.

\section*{References}

\begin{thebibliography}{1}

\bibitem{ZK:PRL:65}
N.~J. Zabusky and M.~D. Kruskal.
\newblock Interaction of {"Solitons"} in a collisionless plasma and the
  recurrence of initial states.
\newblock {\em Phys. Rev. Lett.}, 15:240--243, 1965.

\bibitem{RH:PRL:93}
Philip Rosenau and James~M. Hyman.
\newblock Compactons: Solitons with finite wavelength.
\newblock {\em Phys. Rev. Lett.}, 70:564--567, 1993.

\bibitem{R:PLA:97}
P.~Rosenau.
\newblock On nonanalytic solitary waves formed by a nonlinear dispersion.
\newblock {\em Phys. Lett. A}, 230:305--318, 1997.

\bibitem{CH:PRL:93}
R.~Camassa and D.~Holm.
\newblock An integrable shallow water equation with peaked solitons.
\newblock {\em Phys. Rev. Lett.}, 71:1661--1664, 1993.

\bibitem{FOKAS:PD:95}
A.~S. Fokas.
\newblock On a class of physically important integrable equations.
\newblock {\em Physica D}, 87:145--150, 1995.

\bibitem{GG:JCP:90}
L.~R.~T. Gardner and G.~A. Gardner.
\newblock Solitary waves of the regularised long wave equation.
\newblock {\em J. Comput. Phys.}, 91:441--459, 1990.

\bibitem{LOR:Preprint}
Y.~A. Li, O.~J. Olver, and P.~Rosenau.
\newblock Non-analytic solutions of nonlinear wave models.
\newblock Preprint, 1998.

\end{thebibliography}

\end{document}